\documentclass{article}
\pdfoutput=1
\usepackage{spconf,amsmath,graphicx}
\usepackage{comment}
\usepackage{mathtools}
\usepackage{amssymb}
\usepackage{csquotes}
\usepackage{multirow}
 \usepackage{cite}
\usepackage{comment}
 \usepackage[table,xcdraw]{xcolor}

\title{Health monitoring of Industrial machines using Scene-aware  threshold selection}
\name{Arshdeep Singh$^{1}$, Raju Arvind$^{2}$, Padmanabhan Rajan$^{3}$}
\address{
$^{1,3}$Indian Institute of Technology, Mandi, India, $^{2}$Intel Corporation, Bangalore, India\\
Email: $^{1}$d16006@students.iitmandi.ac.in, $^{2}$raju.arvind@intel.com, $^{3}$padman@iitmandi.ac.in}
\begin{document}
%
\maketitle
\begin{abstract}
This paper presents  an autoencoder based unsupervised approach to identify anomaly in an industrial machine using sounds produced by the machine. The proposed framework is trained  using log-melspectrogram representations of the sound signal. In classification, our hypothesis is that the reconstruction error computed for an abnormal machine is larger than that of the a normal machine, since only normal machine sounds are being used to train the autoencoder. A threshold is chosen to discriminate between normal and abnormal machines. However, the threshold changes as surrounding conditions vary. To select an appropriate threshold irrespective of the  surrounding, we propose a scene classification framework, which can classify the underlying surrounding. Hence, the threshold can be selected adaptively irrespective of the surrounding. The experiment evaluation is performed on MIMII dataset for industrial machines namely fan, pump, valve and slide rail. Our experiment analysis shows that utilizing adaptive threshold, the performance improves significantly as that obtained using the fixed threshold computed for a given surrounding only.
\end{abstract}
\section{Introduction}
\label{sec:intro}

Automated health monitoring of industrial machinery can help in avoiding 
unplanned downtime, increased productivity and reduced maintenance schedules. Acoustic monitoring  of machinery provides advantages such as readily available sensors (microphones), non-intrusive sensing, and ability for omnidirectional sensing. The types of automatic health  monitoring can include vibration sensors \cite{vibration_caesarendra2017review, vib_jin2013anomaly, vibration_2_heyns2012combining,vibration_3_galloway2016diagnosis}, but the microphones are non-intrusive, invariant to humid, temperature conditions and there is no requirement of any balancing techniques as being used by the vibration sensors in order to overcome the misalignment error \cite{isavand2020comparison}.

A challenge in building automatic health monitoring system is the availability of sufficient anomalous\footnote{We use abnormal and anomalous term interchangeably.} samples \cite{scarcity_bull2018active}.  Data-driven approaches
for this purpose have the disadvantage of the unavailability of large-scale public datasets. Although, recently, a few \cite{Purohit2019,koizumi2019toyadmos} releases  publicly the industrial machine dataset.  
Apart from this,  it is very difficult and costly to  generate faults in a normal operating machine.

To overcome the data-scarcity problem, most of the studies employ unsupervised approach by utilizing only normal machine data to identify anomaly. In this regard, the studies \cite{Purohit2019,hendrickx2020general,audio_conte2012ensemble,audio_ono2013anomaly,audio_skoizumi2018unsupervised} employ sound signatures for anomaly detection in various applications. The sound signals have been successfully utilized in many other areas as well, such as sound event detection, sound localization, scene classification etc \cite{aytar2016soundnet,salamon2017deep,mesaros2016tut,mesaros2017dcase}.
In this work, we aim  to build an intelligent system which can discriminate between a normally  and an abnormally operating machine using sounds produced by the machine. 

A typical anomaly identification framework, first computes anomaly scores corresponding to the given machine. Next, an operating point or threshold is chosen to decide whether the score corresponds to the normal machine or the abnormal machine. However, identification of the  threshold and an appropriate selection of the threshold under varying noisy condition is a challenging task.  Empirically, we show that the appropriate threshold changes as the surrounding conditions changes in Section \ref{Sec: performance evaluation}.

In this paper,  an autoencoder-based model is utilized  to monitor health of the machine. We propose a threshold identification formulation by utilizing the distribution of reconstruction error obtained from the autoencoder framework by utilizing normal machine sounds. To overcome the variability of threshold, we propose a convolution neural network (CNN) based scene classification framework, which operates in parallel to the anomaly identification framework. The scene classification framework predicts surrounding and accordingly, chooses an appropriate threshold.  The  scene classification and autoencoder model   do not use anomalous machine data in training at all. Therefore, the proposed framework is an unsupervised and posses awareness of the surrounding as well. The key advantages and major contributions of this paper can be summarized as follows:

\begin{itemize}
    \item An unsupervised health monitoring framework is proposed using sounds produced by the industrial machine. Also, a threshold identification formulation is proposed to discriminate between normal and abnormal machine sounds.

    \item  A CNN-based scene-aware framework is proposed for adaptive selection of the threshold under varying surrounding conditions. 

\end{itemize}

The rest of this paper is organized as follows. 
In Section \ref{sec: proposed}, proposed methodology  is described. Performance evaluation is included in Section \ref{sec: performance eval}. Section \ref{sec: conclusion} concludes the paper.

\section{Proposed Methodology}
\label{sec: proposed}
In this section, first, we explain  feature representation of a sound signal. Next, anomaly identification framework using an autoencoder (AE) is described. Subsequently, a scene classification framework for appropriate selection of threshold is presented in detail.

\subsection{Feature representation of a sound signal}
\label{sec: feature represe}
A given audio recording is converted into a spectrogram using the short time Fourier transform (STFT) at a sampling frequency of 16kHz and a 50\% overlapping window. Next, log-melspectrogram  representations are obtained  with 64-mel bands followed by logarithmic transformation on mel energies. A window with context size of 5 is used to obtain contextual representations. Each contextual representation $\in$ $\mathbb{R}^{320}$ (320= (64 $\times$ 5)), is used as a training instance for the autoencoder model.

\subsection{AE-based anomaly identification}


    
The AE-model comprises of fully-connected layers with 320-64-32-32-64-320 units in the each of the layers. The ReLU activation function is used in all layers. The AE-model has the total number of trainable parameters as approx. 47k. The input to the AE is the contextual representations of 320-dimension as explained previously. The objective function of the AE-model is  to minimize the reconstruction error between training instances and predicted training instances, corresponding to the normal machine sounds as given in Equation \ref{Equ: reconstructed error}. Here, $\Delta_s$ is the reconstruction error between  the contextual representations, $x_i$ and $x_i^p$. $x_i^p$ is predicted instance of $x_i$ using the AE-model and $n$ represents size of $x_i$. 

\begin{equation}
    \label{Equ: reconstructed error}
    \Delta_s= \frac{1}{n}\sum_{i=1}^{n}(x_i-x_i^p)^2 
\end{equation}

\noindent \textbf{Decision making during testing:}
\noindent For a test audio, the contextual representations are obtained as explained in Subsection \ref{sec: feature represe}. Next, the reconstruction error, as given in Equation \ref{Equ: reconstructed error}, is computed for each of the contextual frame using the trained AE-model. Finally, the total reconstruction error corresponding to the test example is computed by averaging the reconstruction error obtained for each contextual frame.  Since, the AE-model is trained using only normal machine sounds, therefore, our hypothesis is that the total reconstruction error for the normal machine is lesser than that of the anomalous machine. A threshold is chosen, which decides the discrimination between normal and abnormal class.

\noindent \textbf{Computation of threshold:} In this work, we propose to compute threshold ($\tau$) as given in Equation \ref{equ: threshold}.

\begin{equation}
    \label{equ: threshold}
    \tau  = \mu(\Delta_v) + \alpha \cdot \sigma(\Delta_v),
\end{equation}

\begin{equation}
    \alpha  = \frac{1}{1 + \frac{\mu(\Delta_v)}{\mu(\Delta_t)}}.
    \label{equ:alpha}
\end{equation}  

\begin{figure*}[t]
    \centering
    \includegraphics[scale=0.55]{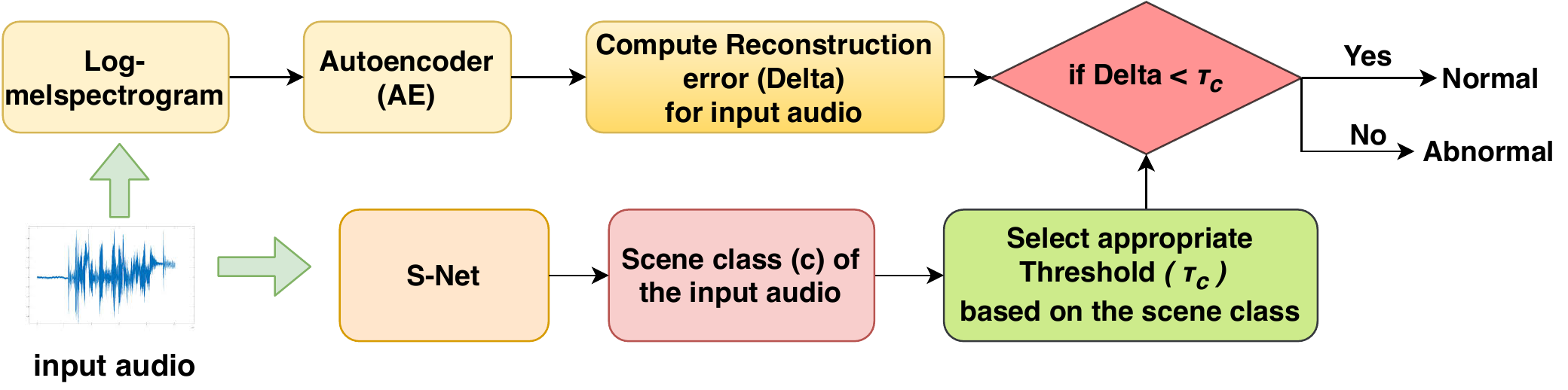}
    \vspace{-0.5cm}
    \caption{Overall evaluation framework for health monitoring of an industrial machine. Here, $c$ denotes the scene class or surrounding predicted by the S-Net. The $\tau_c$ denotes the threshold obtained for the $c^{th}$ surrounding using Equation \ref{equ: threshold}. }
    \label{fig:overall framework}
\end{figure*}

Here, $\mu$ and $\sigma$ represents mean and standard deviation of a set of data respectively. $\Delta_v$ and $\Delta_t$ are sets with elements representing the total reconstruction error corresponding to each example in the validation and training dataset respectively. $\alpha$ $\in [0,1]$, as given in Equation \ref{equ:alpha}, represents a scaling factor for deviation around the mean of the reconstructions error. The proposed threshold ($\tau$) represents the average behaviour of reconstruction error for normal machine sounds. We include a maximum unit deviation margin from the average reconstruction error to handle the model complexity, which may arise due to over-fitting of the autoenconder model on training data. The deviation is further controlled by $\alpha$ parameter depending upon the ratio of average reconstruction error of validation and training dataset from normal machine sounds.

The $\tau$ can vary under different noisy conditions. Using a fixed $\tau$, computed in a given surrounding  can affect performance severely. The threshold under different noisy conditions, denoted as $\tau_c$ (here, $c$ denotes the given surrounding condition), can be computed similarly as given in Equation \ref{equ: threshold}, but using training and validation samples collected in that surrounding. 


\subsection{Adaptive selection of threshold}

To select an appropriate $\tau_c$ in varying surrounding conditions, we propose a 1D-CNN based scene (surrounding) classification model, which is referred to as \textit{S-Net}. The S-Net classifies  surroundings into three noisy levels ($c \in $ \{more noisy, noisy, less noisy\}). The S-Net, thus, can be utilized to select threshold based on the surrounding noisy conditions. In this work, we fix the surrounding into previously mentioned three noisy levels only. In the future, more noisy levels can be used. 

The S-Net comprises of 4-layers. The first layer is convolution layer with 16 filters, each of length 64, followed by a global average pooling layer, dense layer with 64 units and classification layer with 3 units. Each unit has ReLU activation in the network except classification layer.  The total number of parameters of S-Net are approx. 2.4k. The input to the S-Net is a 1-dimensional vector of length $d$. 
The S-Net is trained using raw segments of audio signals corresponding to the various noise levels. An audio recording  is divided into $\mathbf{M}$ smaller non-overlapping segments, $\{x_1,x_2,...,x_{\textbf{M}}\}$. Each segment  $x_i$ $\in$  $\mathbb{R}^d$ is considered as a training instance. 
During testing, the probability scores obtained from each test segments of a given audio, are aggregated together. The output unit corresponding to maximum aggregated score is chosen as the ultimate scene class. Utilizing the scene information corresponding to the input audio, the adaptive threshold can be selected accordingly. The overall proposed evaluation framework is shown in Figure \ref{fig:overall framework}.

\section{Performance Evaluation}
\label{sec: performance eval}


\subsection{Datasets Used}

We utilize MIMII dataset \cite{Purohit2019} to evaluate the proposed framework for health monitoring of industrial machines. The dataset comprises of normal and abnormal sounds from four industrial machines namely (a) fan, (b) pump, (c) valve and (d) slide rail. Each type of machines consists of multiple individual machine models, which are specified by model identity (ID). For example, fan machine has four ID's, ID\_00, ID\_02, ID\_04 and ID\_06. 
Each of the audio signals has 10s length and is recorded at 16kHz sampling rate using 8-microphones.

The dataset comes with machine sounds at three different signal-to-noise ratio (SNRs). A real factory noise is recorded in multiple factories. The noise is added in the original machine sounds to generate audio examples at three different SNR; -6dB, 0dB and 6dB. 
It is important to note that there are three similar set of audio recordings, but, at three different SNR, which are available publicly.

\subsection{Training and validation split}

\textbf{AE-model setup:}
The AE-model is trained for each machine type and ID using 6dB examples of normal sounds of the particular machine type and ID. The training data  consists of 300 audio examples (indexed from 1 to 300, as given in the dataset) from normal sounds of 6dB SNR. From rest of the  normal examples, we choose randomly 300 audio examples (except slide rail ID\_04 and ID\_06, where 100 examples are being used) from each of the SNR dataset as a validation dataset for threshold computation. All other examples (referred to as \enquote{evaluation dataset}) are used to evaluate the proposed framework. 

An audio signal of 10s length is sampled at 16kHz and converted into a single channel by averaging all channels. The log-melspectrogram corresponding to the audio has a size (64 $\times$ 313). This gives a total of 309 contextual frames using a context window of 5 with unit stride as explained in Section \ref{sec: feature represe}. Under these settings, the training dataset consists of 92700-examples ( 92700= 309 $\times$ 300 ) each of 320-dimension. AE-model is trained using Adam optimizer \cite{kingma2014adam} for 5k epochs using mean square error as a loss function.

\noindent\textbf{S-Net setup:}
 S-Net is trained for each machine type and ID using  normal sounds from three scene classes namely; -6dB, 0dB and 6dB SNR. The training set consists of 300 audio examples from each of the SNR dataset. It is important to note that the similar examples as used in training AE-model, are being utilized in training S-Net as well. The S-Net is trained using Adam optimizer for 100 epochs. The loss function is cross-entropy and early stopping is applied with minimum  loss criterion on randomly selected 10\% examples from the training data. An audio signal is downsampled at 8kHz and divided into non-overlapping segments of 250ms segments ($\mathbf{M}$=  40). The total training dataset of all three classes consists of 36k segments (36k= 300 $\times$ 40 $\times$ 3), each of size 2k. 

Area under the curve (AUC), true positive\footnote{Here, positive indicates the abnormal data \label{footnote: postive}} rate (TPR) and false positive rate (FPR) metrics are used for evaluation.

\begin{figure}[htbp]
    \centering
    \includegraphics[scale=0.5]{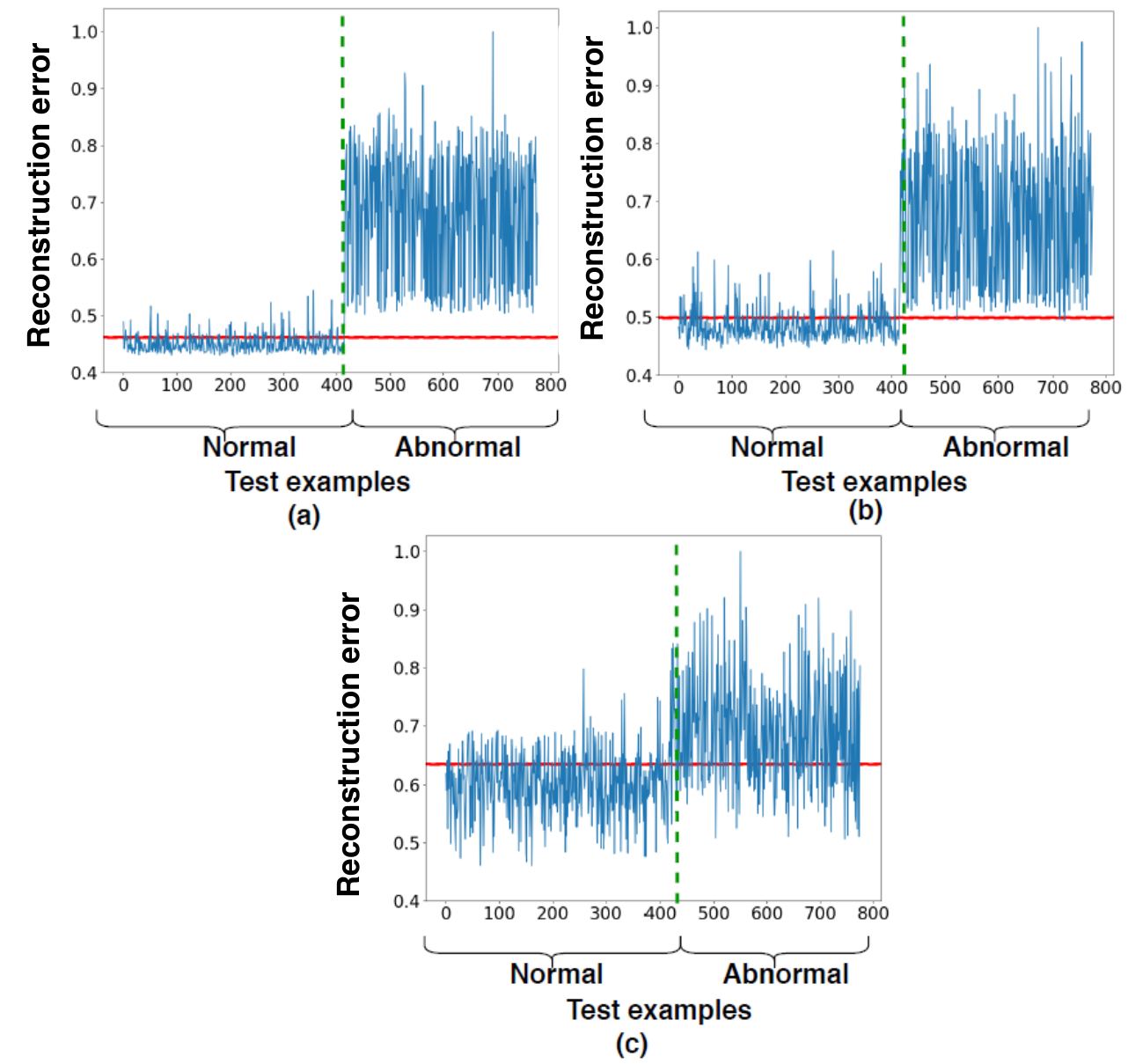}
    \vspace{-0.5cm}
    \caption{ Reconstruction error (normalized) plots for fan ID\_06 (a) 6dB, (b) 0dB and (c) -6dB test examples. The red dotted line shows the threshold obtained for each of the SNR levels using Equation \ref{equ: threshold}.}
    \label{fig: FAN MSE}
\end{figure}

\subsection{Performance Analysis}



\label{Sec: performance evaluation}
The reconstruction error obtained using the AE-model for  evaluation dataset of fan ID\_06 at different SNR dataset is shown in Figure \ref{fig: FAN MSE}. It can be observed that the reconstruction error for the abnormal examples is higher than of the normal examples under all noisy conditions. It shows that the proposed AE-model can be utilized to identify the health (either normal or abnormal) of a machine.

Table \ref{tab: AUC aveergaed} shows AUC for different machine type and ID for different SNR dataset. Mostly, the AUC is significantly greater than 0.5 for various machines. As noise level increases, the AUC decreases for all the machines.

\begin{table}[htbp]
\centering
\caption{Area under the curve (AUC) for various machines averaged across various ID's
at various SNR levels.}
\label{tab: AUC aveergaed}
\resizebox{0.20\textwidth}{!}{%
\begin{tabular}{|c|c|c|c|}
\hline
\multirow{2}{*}{Machine type} & \multicolumn{3}{c|}{AUC} \\ \cline{2-4} 
 & 6dB & 0dB & -6dB \\ \hline
Fan & 0.92 & 0.83 & 0.65 \\ \hline
Pump & 0.86 & 0.82 & 0.73 \\ \hline
Valve & 0.75 & 0.68 & 0.53 \\ \hline
Slide rail & 0.93 & 0.89 & 0.74 \\ \hline
\end{tabular}%
}
\end{table}

\begin{figure}
    \centering
    \includegraphics[scale=0.43]{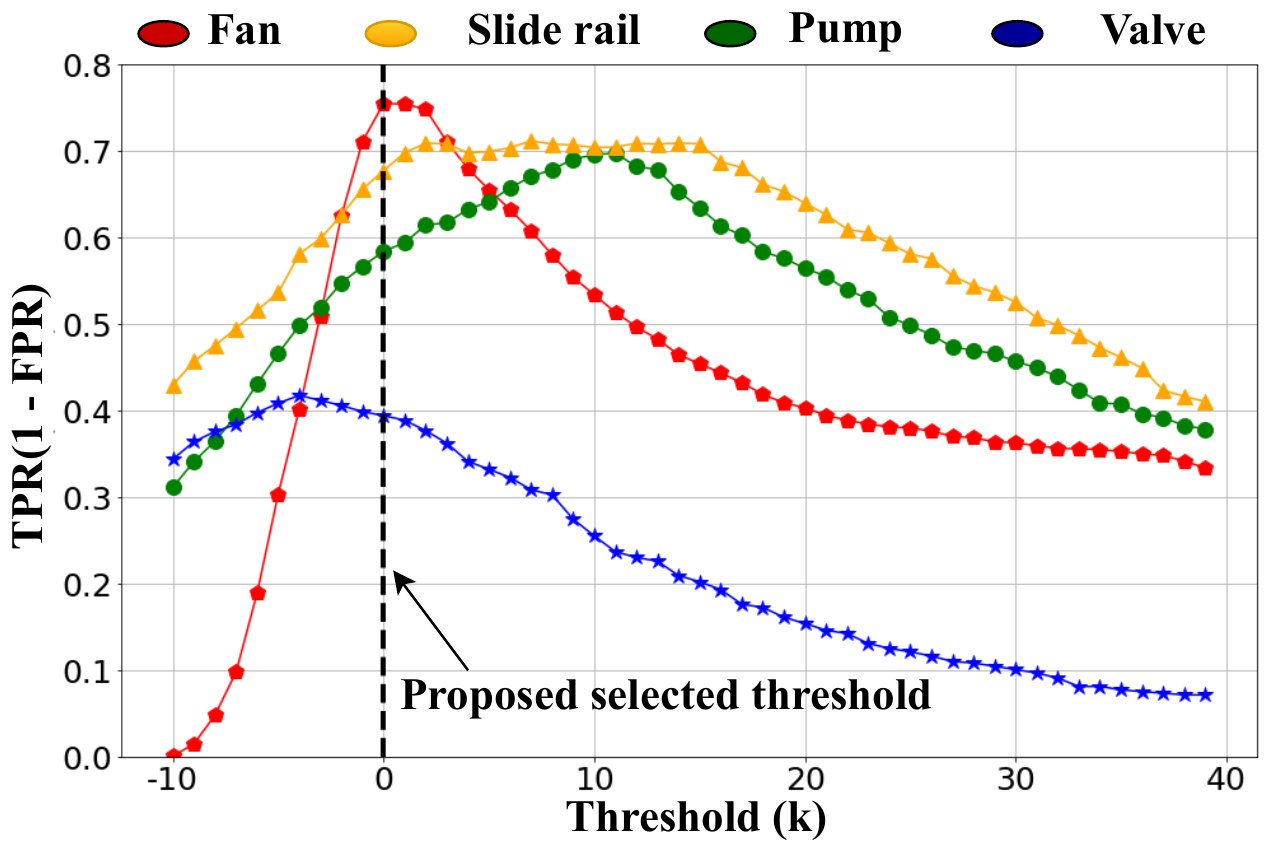}
    \vspace{-0.5cm}
    \caption{ TPR $\times$ (1-FPR) obtained at  different thresholds for  fan, slide rail, pump and valve at 6dB. The various thresholds are computed by varying $k$ in ($\tau + k \times100$). Here, $\tau$ is same as given in Equation \ref{equ: threshold}. The proposed selected threshold is shown at $k=0$.}
    \label{fig: ROC and proposed threshold point}
\end{figure}

Next, the effectiveness of the proposed threshold criterion as given in  Equation \ref{equ: threshold} is measured by obtaining TPR $\times$ (1 -FPR), at randomly selected thresholds. The performance is shown in Figure \ref{fig: ROC and proposed threshold point} for various machines. The performance obtained  for all machines, at the threshold selected using Equation \ref{equ: threshold}, has a maximum absolute deviation of approx. 0.1 as that of obtained at any other threshold.

\begin{figure}[t]
    \includegraphics[scale=0.52]{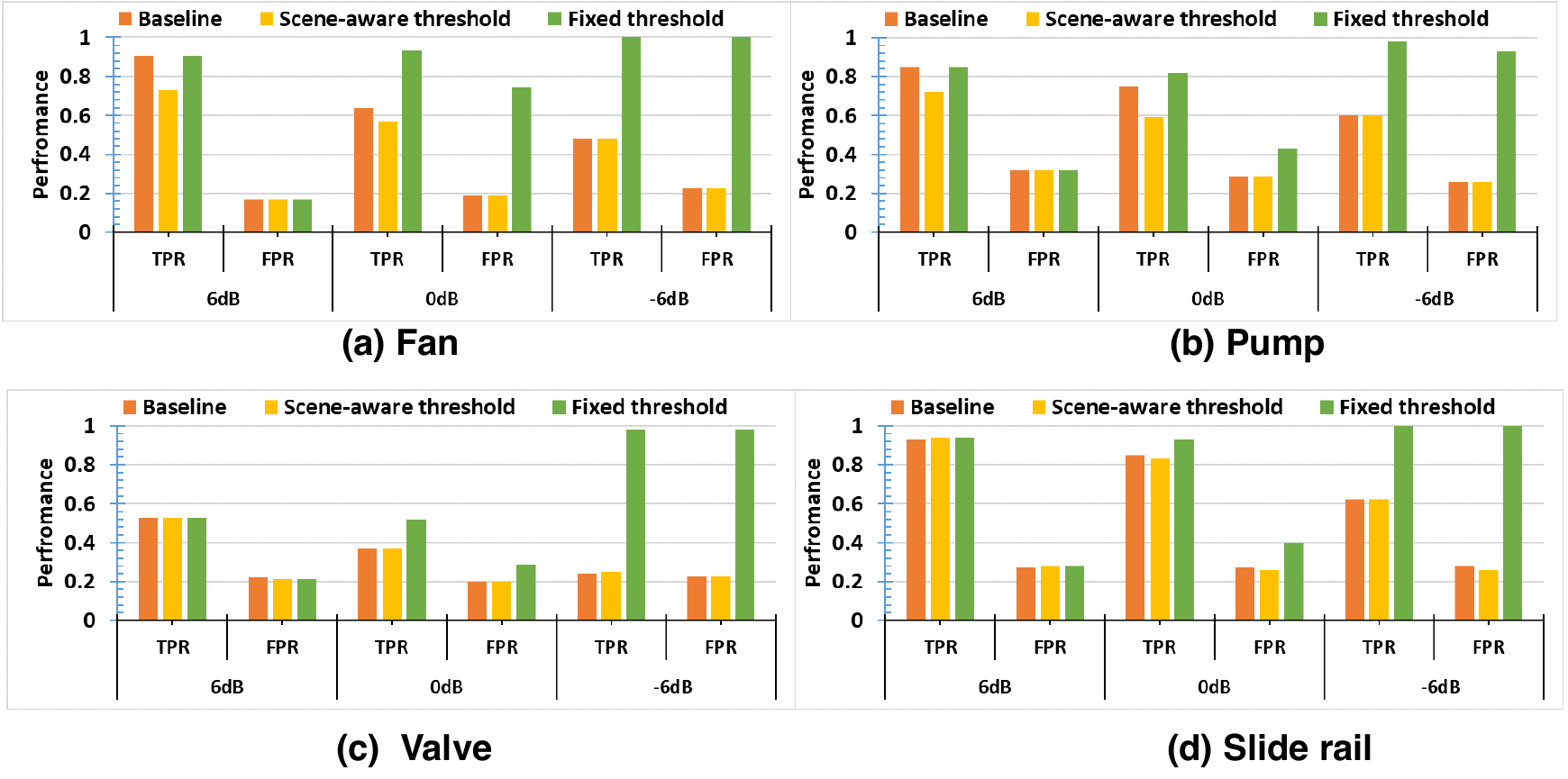}
    \vspace{-0.80cm}
    \caption{Averaged TPR, FPR across various machine ID's for evaluation dataset of (a) fan, (b) pump, (c) valve and (d) slide rail machines at different SNR for baseline, scene-aware threshold and fixed threshold evaluation cases.
    }
    \label{fig: MIMII performance comparison}
\end{figure}

\noindent \textbf{Analysis of performance at different noise levels:} The appropriate threshold computed for different SNR dataset is shown as red dotted line in Figure \ref{fig: FAN MSE}. It can be be observed that the threshold (red dotted line) varies as the surrounding condition changes. This leads to reduce performance drastically, when a fixed threshold, computed under a given surrounding condition, is selected to evaluate the performance under varying surroundings. Figure \ref{fig: MIMII performance comparison} compares TPR, FPR among three  evaluation cases (a)-(c), as explained below, for various machines at different SNR. 

\noindent \textbf{(a) Baseline:} In this case, the performance is computed  for a given SNR dataset using the threshold computed for that SNR dataset only. \\
\noindent \textbf{(b) Scene-aware threshold:} In this case, the threshold is computed for each SNR dataset independently and evaluation is performed by selecting appropriate threshold using S-Net framework as shown in Figure \ref{fig:overall framework}.\\
\noindent \textbf{(c) Fixed threshold:} This case is similar to the  case (b), except that the threshold is computed using 6dB data and is used for evaluation across other SNR dataset. 

The TPR, FPR obtained using scene-aware threshold for most of the machines is similar to that of baseline. However, for fixed threshold, the TPR, FPR across all machines approaches to 1 as the SNR level decreases. This shows that the performance obtained using the fixed threshold biases  towards the  abnormal class only. On the other hand, using scene-aware threshold framework, the performance still remains similar to the baseline performance. This shows the effectiveness of the proposed scene-aware threshold to select appropriate threshold, which can cope up the variations occurred due to the surrounding environment. 

\section{Conclusion}
\label{sec: conclusion}
In this paper, we propose an unsupervised health monitoring framework, which identifies and  adaptively selects an appropriate threshold, to identify the anomaly using sounds produced by the industrial machine. The proposed health monitoring framework requires only normal machine sounds and surrounding conditions, which are easy to collect. Therefore, the proposed approach can be easily utilized in the real-factory to monitor health of industrial machines. In future, we aim to utilize more SNR levels to validate the effectiveness of the proposed framework.
\bibliographystyle{IEEE}
\bibliography{ref}

\end{document}